\documentclass[aps,prd,nofootinbib]{revtex4}
\usepackage{latexsym}
\usepackage{graphicx}
\usepackage{amsmath,amsfonts}
\usepackage{amsbsy}
\usepackage{mathrsfs}
\usepackage{color}
\usepackage{psfrag}
\usepackage{enumerate}
\usepackage{amsmath,amssymb,calc,amsfonts}
\usepackage{latexsym}
\DeclareFontFamily{U}{rsfs}{}         
\DeclareFontShape{U}{rsfs}{m}{n}{<5> rsfs5 <6><7> rsfs7          %
  <8><9><10><10.95><12><14.4><17.28><20.74><24.88> rsfs10}{}     %
\DeclareMathAlphabet{\mathfs}{U}{rsfs}{m}{n}                     %
\definecolor{indiagreen}{rgb}{0.07, 0.53, 0.03}
\def\beq{\begin{eqnarray}}
\def\eeq{\end{eqnarray}}
\def\l{\ell}
\def\mb#1{\mathbb{#1}}

\def\nn{\nonumber\\}

\def\mb#1{\mathbb{#1}}

\def\={\stackrel{\Delta}{=}}

\def\mbf#1{\mbox{\boldmath${#1}$}}

\def\lie{\pounds}

\def\half{{\textstyle{\frac{1}{2}}}}

\begin{document}

\title{Effective Quantum Theory of Black Hole Horizons}

\author{Ayan Chatterjee}\email{ayan.theory@gmail.com}
\affiliation{Department of Physics and Astronomical Science, Central University 
of Himachal Pradesh, Dharamshala -176215, India.}
\author{Amit Ghosh}\email{amit.ghosh@saha.ac.in}
\affiliation{Theory Division, Saha Institute of Nuclear Physics, 1/AF Bidhan 
Nagar, Kolkata 700064, INDIA.}
\begin{abstract}
In this paper, we  develop an effective quantum theory of black
hole horizons using only the local horizon geometry. 
On the covariant phase space of the Holst action admitting Weak Isolated Horizon as an inner boundary,
we construct Hamiltonian charges corresponding to
Lorentz symmetries. We show that horizon area is the Hamiltonian charge 
corresponding to Lorentz boosts as well as that for Lorentz rotation which acts on $2$-sphere cross- sections of the 
horizon. Using this expression of area as a generator of Lorentz rotation, and the fact that
quantum states residing on the horizon cross- sections carry a representation of $ISO(2)$,
we derive the spectrum of area operator on the horizon.
The eigenstates of this area operator are shown to be labelled by integers or half integers. The entropy is obtained completely
in terms of these \emph{area quanta} residing on the horizon,
and is shown to have exponentially suppressing corrections to the area law. The formalism is also extended 
to non- minimally coupled scalar fields, where the area operator gets modified due to the 
value of the scalar field on the horizon. 
\end{abstract}
\maketitle
\section{Introduction}
\label{intro}
The classical dynamics of black hole horizons encoded
in the laws of black hole mechanics points to its thermal nature
\cite{Bardeen:1973gs,Hawking:1971vc,Hawking:1974sw,Bekenstein:1973ur,Bekenstein:1974ax}.
It has now become a well established fact that the description 
of gravity as spacetime dynamics can explain the thermodynamic nature of horizons.
Indeed, there exist deep clues relating thermodynamical quantities with gravitational or geometric quantities
on the horizon. More precisely, it has been shown that black holes have a temperature $T=\hbar\kappa/2\pi$, and 
the first law of black hole mechanics requires that black holes of area 
$\mathcal{A}$ must have thermodynamic entropy given by $S=(\mathcal{A}/4\ell_{p}^{2})$ 
\cite{Hawking:1974sw,Bekenstein:1973ur,Bekenstein:1974ax,Wald:1995yp,Wald:1993nt,Iyer:1994ys},
where $\ell_{p}$ is the Planck length. 
There have been several attempts to understand the microscopic origin of 
this black hole entropy. In particular, in string theory as well as in loop quantum gravity, the 
microstate counting not only gives the Bekenstein- Hawking area law, but also
successfully generates the entropy corrections beyond the logarithmic terms. It is possible
that both these theories actually describe the same physics but are using different variables
or descriptions. However, in absence of consensus on the correctness of either theories, it is natural
to look for alternative representations of black hole which leads to similar interrelations between 
the classical and quantum nature of black hole horizons using the local geometrical structures 
of the horizon itself without any reference the asymptotic structures.

A very useful classical notion of black hole which is not interacting
with its surroundings, (in other words, no matter is falling through it), is 
provided by the Weak Isolated Horizon (WIH) formalism \cite{Ashtekar:1998sp, Ashtekar:2000sz, Ashtekar:2000hw, Ashtekar:2004cn}. 
In this set- up, an isolated black hole horizon, in $4$- dimensional spacetime, 
is described as a \emph{non- expanding} $3$- dimensional
null surface foliated by marginally trapped $2$- spheres. More precisely, one assumes
the null normal to the horizon $\ell^{a}$ is expansion free, $\theta_{(\ell)}=0$,
shear- free $\sigma_{(\ell)}=0$ and twist- free,
and the field equations of matter and geometrical fields hold good on the horizon. 
A non- expanding horizon (NEH) is called a WIH if the connection on the normal bundle is also
lie dragged on the horizon. 
A NEH is a good characterisation of a black hole horizon.
and it is natural to view the horizon as an inner boundary of the spacetime in this framework. 
In this formalism, one may envisage situations where the horizon and the fields on it are
time independent but the near horizon spacetime is highly dynamical comprising
of matter fields as well as gravitational and electromagnetic radiation. In such a scenario too, it is 
possible to use the WIH formalism to construct quasilocal quantities like mass $M$, angular momentum $J$, and surface
gravity $\kappa$ without any reference to the asymptotic infinity, and thereby prove the zeroth and the first law of black hole
mechanics in a purely quasilocal setting (quasilocal refers to the fact that the definition
requires the cross-section of the horizon and a finite element of the horizon, that is, a point on the horizon and its finite neighbourhood).
The WIH formalism has been used
to obtain the black hole entropy in the setting of quantum geometry. The horizon is characterised quantum mechanically
by Chern- Simons theory residing on punctured two spheres, foliating the quantum horizon
\citep{Ashtekar:2004cn, Ashtekar:1999wa, Ashtekar:1997yu, Ashtekar:2000eq}. Naturally, black hole entropy
is obtained considering all the quantum states of this horizon geometry. For large 
black holes of fixed area, logarithm of the number of states, and hence the entropy, is precisely the area. Further corrections to the
area law can also be obtained by carefully counting all the states upto the requisite order
\citep{Kaul:2000kf, Meissner:2004ju, Domagala:2004jt, Ghosh:2004wq}. So, the isolated horizon
formalism has broadened the applicability of the laws of black hole mechanics to a large class of black hole horizons in equilibrium and
has introduced a set of highly restrictive boundary conditions leading to the classical and quantum description of quasilocal horizons.

However, there are some drawbacks in this abovementioned derivation of entropy of the black holes.
It relies on the spectrum of the area operator as an external input
which may not correctly represent the area spectrum on a black hole horizon. The area operator and its
spectrum has been derived for a $2$-dimensional submanifold of a spacelike surface ${M}$ in the spacetime \cite{Ashtekar:2004eh}.
It is not clear if the area spectrum for a $2$- dimensional submanifold of a null surface ${\Delta}$ should give 
identical result (in fact there have been claims of a very different area spectrum for Schwarzschild like black holes \cite{Bekenstein_Mukhanov}). 
Furthermore, the quantum description of horizon requires a well-defined compatibility condition between 
the bulk and the boundary Hilbert spaces. This requirement forces the $2$-sphere
cross sections to be punctured by bulk links which \emph{deposit} area elements on this sphere \cite{Ashtekar:2000eq}.
Thus, the quantum theory requires a well- defined interaction between the bulk and the boundary.
To remedy this, one needs an expression for area of the horizon cross-section directly from a 
classical description of black hole horizon and then, relate it,
in some way to a quantum description and eventually to a thermodynamical relation on entropy. 
The usual classical laws of black hole mechanics are not helpful in reproducing an expression for area since they involve variations of 
horizon area, whereas we need an expression for the horizon area itself.
One attempt to derive a local relation between geometry and thermodynamics on the horizon was 
given in \cite{Frodden:2011eb}. Using the physical process version of the first law of black hole mechanics,
where one considers the change in the black hole parameters due to absorption of infalling test particles,
it was  argued that observers fixed at a proper distance $l_{0}$, very close from the horizon
of a stationary black hole may define a notion of \emph{energy} ($E$) proportional to the \emph{horizon area} ($\mathcal{A}$),
given by $E=A/8\pi G l_{0}$. However, this relation also has a drawback that it is observer dependent and
requires near horizon geometry (see also \cite{Bianchi:2012vp, Bianchi:2012ui, Chatterjee:2015lwa}). Ideally, one would like to derive the area spectrum of the horizon without using geometrical structures of the bulk spacetime. Since the WIH formalism 
is a local description of black holes, it should be our best possible choice
to evaluate a classical expression of the horizon area.  

The main crux of this paper is to derive a classical expression of the horizon area, directly from the WIH formalism, and utilize it to 
understand the microscopic description of black hole entropy. In this formalism, quantum states responsible for entropy
will be the ones residing purely on horizon only and no connection with the bulk
will be needed. In this sense, and unlike the framework of \cite{Ashtekar:2000eq},
this horizon may be thought to be classically as well as quantum mechanically \emph{isolated}
from the environment. Along the way to these formulation,  several new results have been
proved and these are arranged in the following manner.

The next section (section II) contains a discussion of geometry 
of WIH formalism, the symmetries of bulk spacetime and the residual symmetries on WIH boundary.
The residual symmetries are those transformations which preserve the WIH boundary conditions. Most of these were 
discussed earlier in \cite{Chatterjee_ghosh_basu}. The third section contains the Holst symplectic current and
construction of the symplectic structure. Section (IV)
has derivation of the Hamiltonian charges corresponding to generators of the symmetry vector fields on WIH. 
We show that, on this phase- space of WIHs, 
horizon area is the generator of two transformations: a Lorentz boost along the 
horizon, and a Lorentz rotation on the $2$- sphere cross- sections of the horizon.
Additionally, using these two results, we establish the \emph{simplicity constraint} \cite{Rovelli:2013osa}) on the null surface of WIH. 
Note that it has been argued earlier too that the horizon area is canonically conjugate to the horizon boost,
and although some arguments and proof exist, they refer to the near horizon structure
of a non- extremal black hole \cite{Carlip:1993sa,Massar:1999wg,Wall:2010cj,Wall:2011hj, Chatterjee:2015lwa}. The proof presented here
uses geometrical structure on the horizon only and does not refer to the bulk spacetime. Furthermore,
this derivation, since it only deals with the horizon generator, works quite naturally for extremal black holes as well.
The new additional finding is that the horizon area (modified appropriately by the Immirzi
parameter) is also the generator of the Lorentz rotation of the $2$-sphere
cross-sections of the horizon. The proof of simplicity constraint developed
here is also crucial since it shows that the connection arises naturally for shear- free and expansion- free null hypersurfaces. 
The fifth section extends the calculation to gravity theories with non- minimally coupled scalar fields
using a non- minimally coupled Holst action.
Our analysis on non- minimally coupled Holst action gives two further results: the generator of 
boost is the horizon area modified by the value of the scalar field on the horizon, and that 
the horizon area (modified appropriately by the Immirzi
parameter as well as the value of the scalar field on the horizon) the generator of Lorentz rotation of the $2$- sphere
cross- sections of horizon. We also prove the validity of the simplicity constraint for non- minimal scalar field
couplings to gravity and show that it holds 
good even in presence of a non- minimally coupled scalar field.
This expression has been obtained previously too \cite{Ashtekar:2004eh,Ashtekar:2000eq}, but
using a boundary Chern- Simons theory. Here, it arises directly as a Hamiltonian charge. 
Section (VI) gives the area spectrum and a derivation of the black hole entropy.
We obtain the spectrum of area operator by raising the classical expression of area to quantum level,
by acting it as an operator on a well defined Hilbert space of states. 
This is carried out as follows:
On this WIH phase- space, we determine the algebra of charges and show that this algebra
is identical to the $ISO(2)\ltimes \mathbb{R}$ sub- algebra of the Lorentz algebra. The quantum states residing on $2$- sphere
cross- sections of the horizon belong to the representation of the $iso(2)$ and are labelled by integers (or half- integers) (see the Appendix $C$),
and the area acts as an operator on these states. These states are also the eigenstates of area operator and therefore
the area spectrum becomes equidistant. This notion of equidistant area spectrum is not new, and in several of the papers
on black hole spectroscopy including the quasinormal modes of black holes,
this kind of equidistant spectrum has been motivated \cite{Bekenstein_Mukhanov, hod, dreyer, Polychronakos_1}. Such a spectrum 
has also been shown to arise in the context of Quantum Geometry, by using different regularisations of the area operator \cite{Alekseev}.
Our derivation also points to such a equidistant spectrum of the horizon area. In this section,
we also derive the entropy. Our counting reveals that the entropy is exactly equal to the Bekenstein- Hawking result
but admits corrections which are exponentially suppressed. This is a new result and may contain seed of non- perturbative 
corrections to black hole entropy.

Of the results derived here, some of them are completely new and, to our knowledge,
have not been discussed earlier. Few of the results here have been put on a firmer footing
because, all our calculations are based on the WIH formalism and hence,
utilize the local horizon geometry only without relying on the geometrical structure 
of the bulk spacetime. Hence, these results shall appeal to a wider class of black holes,
than those discussed in the literature earlier.


\section{Symmetries on a non- expanding horizon}
Let $\cal M$ be a $4$- dimensional manifold with a metric $g_{ab}$ having signature $(-,+,+,+)$.
In $\cal M$, let ${\Delta}$ be a null hypersurface generated by a future directed null vector field $\ell^a$ 
and foliated by $2$-spheres. Let 
us fix particular a cross- section $S_{0}$ of $\Delta$ with coordinates $(\theta, \phi)$. 
Let $\lambda$ be the affine parameter on $\Delta$ with $S_{0}$ being at $\lambda=0$.
The tangent vector field is  then given by 
$\ell^{a}=(\partial/\partial \lambda)^{a}$. We may 
use the affine parameter $\lambda$ to label the cross- sections on $\Delta$. 
Thus, if $P$ is any point on $S_{\lambda}$, it's coordinates are $(\lambda, \theta, \phi)$, where 
$\lambda$ is the affine separation of the point $P$ from $S_{0}$. For our purpose,
we shall use another parameter $v$ as horizon generating, where $v$ is
related to the affine parameter $\lambda$ through $\lambda=a\,e^{\kappa\, v}+b$.
The horizon generating vector field is then $\ell^{a}=(\partial/\partial v)^{a}$. 
On $\Delta$, one may define a metric induced from the full spacetime. That metric
is degenerate, and one may however define an inverse metric $q^{ab}q_{ac}q_{bd}:=q_{cd}$
which captures essentially the foliation geometry. Since $\Delta$ is null, it is also twist- free.
Also, since the generators $\l^{a}$ are null,
the parallel transport of $\l^{a}$ is also proportional to $l^{a}$. Then, $\l^{a}\nabla_{a}\l^{b}=\kappa_{(\l)}\l^{b}$,
where $\kappa_{(\l)}$ is the acceleration corresponding to the null normal $\l^{a}$ and $\nabla_a$ is the covariant
derivative compatible with $g_{ab}$.
The expansion $\theta_{(\l\,)}$ of the null normal $l^a$ is defined by
$\theta_{(\l\,)}=q^{ab}\nabla_a \l_b$. For convenience, we shall 
use the null tetrad $(\l^{a}, n^{a}, m^{a}, \bar{m}^{a})$ such that $1\!=\!-n_{a}\l^{a}=\! m_{a}\bar m^{a}$ and all other scalar
products vanish. This basis is especially suited for the setup since one of the
null normals $l^a$ matches with one of the basis vector. In this basis the spacetime metric is 
given by $g_{ab}=-2\l_{(a}n_{b)}+ 2 m_{(a} \bar m_{b)}$.

The surface $\Delta$, equipped with the class $[\l^a]$ of null normals ($\ell^{a} \sim \ell^{\prime a}$ if $\ell^{\prime a}=c\ell^{a}$, $c$ being
a constant), is called a \textit{non- expanding horizon} (NEH) in $({\cal M},g_{ab})$ if the following conditions hold 
for all vectors in the equivalence class \cite{Ashtekar:2000hw}:
\begin{enumerate}\label{bcond}
\item $\Delta$ is topologically $S^2 \times \mathbb{R}$.
\item The expansion $\theta_{(\l)}=0$.
\item The equations of motion hold on $\Delta$ and the vector field $-T^a{}_b \l^b$
is future directed and causal on $\Delta$.
\end{enumerate}
The first condition describes the foliation while the second condition, that the horizons be exansion free,  is a crucial
condition applicable to black hole horizons. This condition implies the existence of a well defined and unique connection on $\Delta$, from
the full spacetime connection. The third condition ensures that
equations of motion and energy condition hold.
Along with the Raychaudhuri equation and the energy conditions, the null surface $\Delta$ may be shown 
to be shear- free. These restrictions also imply the existence of a Killing vector field $\xi \l^{a}$ on $\Delta$. 
It should also be said that the NEH is called a WIH if the connection one form on the normal bundle
$\omega^{(\l)}$ is lie dragged $\lie_{l}\omega^{(\l)}=0$. This condition
leads to the constancy of surface gravity $\kappa_{(\l)}$ on the horizon.
Note that all these boundary conditions are intrinsic to $\Delta$. 


Since we shall be using the first order tetrad- connection formalism, the gravitational degrees of freedom
is encoded in the co- tetrads $e_{a}^{I}$ and the gravitational connection one form $A_{IJ}$. 
The quantity $a,b, \dots$ refer to spacetime indices while 
$I,J, \dots$ will be used for internal flat spacetime.
The internal metric $\eta_{IJ}$ is mapped to the spacetime metric $g_{ab}$
through the tetrads, $g_{ab}=e^{I}_{a}\,e^{J}_{b}\,\eta_{IJ}$. For 
any internal vector $\lambda^{I}$, the gravitational 
connection $(A_{aI}{}^{J})$ is obtained through the action of the derivative operator,
$\nabla_{a}\lambda^{I}=\partial_{a}\lambda^{I}+A_{aI}{}^{J}\lambda_{J}$, where $\partial_{a}$ is the 
internal flat connection. 

Given a fixed tetrad $e_{a}^{I}$, one may construct the bulk spacetime metric $g_{ab}=e^{I}_{a}\,e^{J}_{b}\,\eta_{IJ}$,
as well as the inner boundary $\Delta$, with a null normal $\ell_{a}=e_{a}^{I}\,\ell_{I}$ belonging to the equivalence class
of null normals satisfying the boundary condition of a WIH.  The horizon $\Delta$ will also be assumed to have a fixed set 
of internal tetrad basis $(\l^{I}, n^{I}, m^{I}, \bar{m}^{I})$, where, for example,  $\ell^{I}=e_{a}^{I}\ell_{a}$. 
These internal null basis are fixed such that they basis are annihilated by
the internal flat connection (for example, $\partial_{a}\ell^{I}=0$). Note that tetrad  $e_{a}^{I}$ is not unique since they may be modified 
by Lorentz transformations. In the bulk, the spacetime $\mathcal{M}$ allows all the possible Lorentz transformations of the
tetrad $e_{a}^{I}$ since they all shall give the same spacetime metric. However, on $\Delta$, the set of Lorentz transformations are restricted.
All Lorentz transformations are not allowed. Out of all possible $SL(2,C)$ transformations in the bulk, only those are viable 
on horizon which preserve the boundary conditions on $\Delta$ as mentioned above. 
More precisely, given a set of tetrads in the bulk (related by
Lorentz transformations), only those are acceptable which, acting of the fixed $\ell^{I}$, generate $\ell^{a}$s in the equivalence class of WIHs. 
Thus, since Lorentz transformations affect tetrads, they also affect the null vector fields $(\ell^{a}, n^{a}, m^{a}, \bar{m}^{a})$.
For example, a Lorentz transformation changes $e^{a}_{I}$ to $e^{\prime a}_{I}=\Lambda^{J}{}_{I}e^{a}_{J}$ and hence,
on the horizon, where $\ell^{I}$ is fixed, this affects the null normal $\ell^{a}$. The set of possible Lorentz transformations are the ones which preserve the Newman- Penrose coefficients on the horizon $\Delta$, or atleast transform them homogeneously. These transformations
shall be designated as the symmetry of the WIH.

This exercise of determining the symmetries was already carried out in \cite{Chatterjee_ghosh_basu}. Let us briefly recall
the basic arguments. The set of all Lorentz transformations consists of the following set: a Lorentz transformation generating a boost in the
$(\ell-n)$ plane, a rotation in the $(m-\bar{m})$ plane and a mixture of a boost and a rotation
in the $(\ell-m)$ (keeping $n$ fixed) and one further boost- rotation in
the $(n-m)$ plane (keeping $\ell$ fixed). According to the arguments of the previous section, they may also
be viewed as a transformation acting of the spacetime null tetrad and are given as follows:
\begin{eqnarray}
&\ell^{a}\mapsto\xi \ell^{a},n^{a}\mapsto\xi^{-1}n^{a},m^{a}\mapsto m^{a},\label{lor1}\\
&\ell^{a}\mapsto \ell,n^{a}\mapsto n^{a},m^{a}\mapsto e^{i\theta}m^{a},\label{lor2}\\
&\ell^{a}\mapsto \ell^{a},n^{a}\mapsto n^{a}-cm^{a}-\bar c\bar m^{a}+c\bar c\ell^{a},m^{a}\mapsto m^{a}-\bar
c\ell^{a},\label{lor3}\\
&\ell^{a}\mapsto \ell^{a}-bm^{a}-\bar b\bar m^{a}+b\bar b\ell^{a},n^{a}\mapsto n^{a},m^{a}\mapsto m^{a}-\bar
bn^{a},\label{lor4}
\end{eqnarray} 
where $\xi,\theta,c,b$ are smooth functions on $\Delta$. The functions $\xi$ and $\theta$ are real while $c$ and $b$ are complex.
This accounts for the six parameters of the Lorentz transformations. Since a WIH is a expansion- free, twist- free null surface, many of the 
Newmann- Penrose scalars like $\kappa_{NP}$, $\rho$, $\sigma$ vanish on the horizon $\Delta$. So, out of all
these transformations, we should only look for the ones which map the boundary conditions of a WIH to itself.
Under the transformations (\ref{lor1}),
(\ref{lor2}) and (\ref{lor3}), $\kappa_{\rm NP},\rho,\sigma$
transform homogeneously and hence, if they vanish, all horizons generated by these transformed set of null tetrads will also be a WIH.
\begin{align} &\kappa_{\rm NP}\mapsto\xi^2\kappa_{\rm NP},\;\rho\mapsto\xi\rho,\;\sigma\mapsto\xi\sigma\\ 
&\kappa_{\rm NP}\mapsto e^{i\theta}\kappa_{\rm NP},\;\rho\mapsto\rho,\;\sigma\mapsto e^{2i\theta} \sigma\\ 
&\kappa_{\rm NP}\mapsto\kappa_{\rm NP},\;\rho\mapsto\rho-c\,\kappa_{\rm NP},\;\sigma\mapsto\sigma-\bar c\,\kappa_{\rm NP}.\end{align}
However, under (\ref{lor4}) the Newman- Penrose  coefficients 
transform inhomogeneously and to preserve the WIH boundary conditions, $b$ must vanish. This essentially
reduces the set of possible Lorentz transformations.
The Lorentz matrices associated with the transformations
(\ref{lor1})-(\ref{lor3}) are respectively
\begin{align} \Lambda_{IJ}=&-\xi\l_In_J-\xi^{-1}n_I\l_J+2m_{(I}\bar
m_{J)},\label{L1}\\
\Lambda_{IJ}=&-2\l_{(I}n_{J)}+(e^{i\theta}m_I\bar m_J+c.c.),\label{L2}\\
\Lambda_{IJ}=&-\l_In_J-(n_I-cm_I-\bar cm_I+|c|^2\l_I)\l_J +(m_I-\bar c\l_I)\bar m_J+(\bar m_I-c\l_I)m_J .\label{L3}\end{align}
The generators corresponding to these transformations are given by the following quantities:
\begin{align} 
&B_{IJ}=(\partial\Lambda_{IJ}/\partial\xi)_{\xi=1}=-2\l_{[I}n_{J]},
\label{lbb}\\
&R_{IJ}=(\partial\Lambda_{IJ}/\partial\theta)_{\theta=0}=2im_{[I}\bar
m_{J]},\label{lbr}\\
&P_{IJ}=(\partial\Lambda_{IJ}/\partial{\rm Re}\,c)_{c=0}=2m_{[I}\l_{J]}+2\bar
m_{[I}\l_{J]},\label{lbp}\\
&Q_{IJ}=(\partial\Lambda_{IJ}/\partial{\rm Im}\,c)_{c=0}=2im_{[I}\l_{J]}-2i\bar
m_{[I}\l_{J]},\label{lbq}\end{align}
where $B,R$ generate (\ref{lor1}) and (\ref{lor2}) respectively and $P,Q$
generate (\ref{lor3}). A straightforward calculation gives their Lie brackets
\begin{align} &[R,B]=0,\quad [R,P]=Q,\quad [R,Q]=-P,\nonumber\notag\\
&[B,P]=P,\quad [B,Q]=Q,\quad [P,Q]=0,\label{iso2}\end{align}
where $[R,B]_{IJ}=R_{IK}B^K{}_J-B_{IK}R^K{}_J$ and so on. This is the Lie
algebra of $ISO(2)\ltimes\mb R$ where the symbol $\ltimes$ stands for the semidirect
product. $R,P$ and $Q$ generate $ISO(2)$: $R$ generating Euclidean rotations
in the $(m-\bar{m})$ plane, $P$ generates rotation in the $(\ell-{m})$,
$Q$ generates rotation in the $(\ell-\bar{m})$; while, $B$ generates $\mb R$, which are scaling transformations of  $(\ell-{n})$. 
It is not surprising that the NEH boundary conditions are invariant only under this subgroup of
local Lorentz group, since this is the little group of the Lorentz group which keeps the
horizon generator invariant. Now, given a horizon, $\Delta$ is generated by a set of null generators in 
the equivalence class of null normals $[\ell^a]$ and a given set of foliation vector fields. Naturally,
this elevates the group of local rescaling $\mb R$  and, the rotation subgroup in $ISO(2)$ 
generated by $R$, from local to global transformations. In the following, we would like to construct the set of charges
which are Hamiltonians corresponding to these transformations. We shall show that these charges,
which generate boost and angular momentum on the phase- space respectively,
are related to horizon area.

\section{Holst Action and the Space of Solutions}
Let us briefly recall the construction of the covariant phase- space.
Note that since the internal boundary is an isolated horizon, the space of solutions shall admit 
isolated horizon as an internal boundary. To construct the phase- space, we work with the first order Holst Lagrangian and
the covariant phase space formalism. The method is
as follows: Given a Lagrangian $L$, using the equations of motion, the variation
of the Lagrangian is $\delta  L=d\Theta(\delta)$ 
where $\Theta(\delta)$ is called the symplectic potential. The symplectic potential
is a $3$-form in space-time and a 
$0$-form in phase space. This symplectic potential gives the symplectic current
$\mathcal{J}(\delta_1,\delta_2)= \delta_1\Theta(\delta_2)-\delta_2\Theta(\delta_1)$, 
which, by definition, is closed on-shell.  Note that the the symplectic current
is essentially the on- shell second variation of the Lagrangian. The symplectic structure is obtained from this
symplectic current:
\begin{equation}
\Omega(\delta_1,\delta_2)=\int_{M}\mathcal{J}(\delta_1,\delta_2)
\end{equation}
where $M$ is a space-like hypersurface. Note that if the equations of motion and linearized equations of motion hold,  
the symplectic current is conserved $d\mathcal{J}=0$. This current conservation equation,
when integrated over a closed region of spacetime bounded by $M_+\cup M_-\cup 
\Delta$ (where $\Delta$  is the inner boundary considered) gives:
\begin{equation}
\int_{M_+}\mathcal{J}-\int_{M_-}\mathcal{J}~+~\int_{\Delta}\mathcal{J}=0,
\end{equation}
where $M_+,M_-$ are the initial and the final space-like slices, respectively. For the case when WIH is an internal boundary, third term becomes
exact, $\int_{\Delta}\mathcal{J}=\int_{\Delta}d\bar{\mathcal{J}} $, and the hypersurface independent symplectic structure is given by:
\begin{equation}
\Omega(\delta_{1}, \,\delta_{2})= \int_M \mathcal{J}(\delta_1,\delta_2)-\int_{S_\Delta}\bar{\mathcal{J}}(\delta_1,\delta_2)
\end{equation}
where $S_\Delta$ is the 2-surface at the intersection of the hypersurface  $M$ with 
the boundary $\Delta$. 
The quantity $\bar{\mathcal{J}}(\delta_1,\delta_2)$ is called the boundary symplectic current.


\begin{figure}
  \centering
    \includegraphics[width=0.5\textwidth]{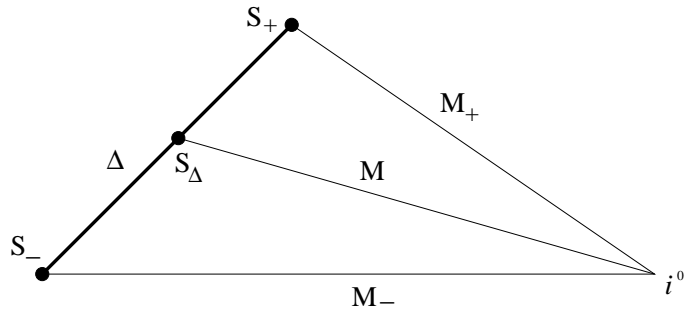}
    \caption{The line $\Delta$ is the weak isolated horizon, $M_{+}$ and $M_{-}$
    are the two Cauchy surfaces which meet $\Delta$ on the cross- sections $S_{+}$ and $S_{-}$ 
    respectively. }
\end{figure}

Here, we shall use the first order formalism in terms of tetrads and connections. 
This formalism is naturally adapted to the WIH formalism since, 
that the boundary conditions are easier to implement and the construction of the covariant phase- space becomes simpler.
Furthermore, the first order formalism separates the action of local Lorentz transformations and spacetime
diffeomorphisms, and since we are mainly interested in the Hamiltonian charges of Lorentz transformations,
the first order formalism is suited. For the first order theory, we take the fields on the manifold to be
($e_{a}{}^{I},\, A_{aI}{}^{J}$), where $e_{a}{}^{I}$ is the co- tetrad, $A_{aI}{}^{J}$
is the gravitational connection.  The Holst action in first order gravity is given by the following Lagrangian (the factor $16\pi G\gamma$ is a constant) \cite{Holst:1995pc}:
\begin{equation}\label{lagrangian1}
-16\pi G\gamma~L= \gamma \Sigma_{IJ}\wedge F^{IJ}~-~e_{I}\wedge e_{J}\wedge F^{IJ}~ -~\gamma~d(\Sigma_{IJ}\wedge A^{IJ}) +~d(e_{I}\wedge e_{J}\wedge A^{IJ}),
\end{equation}
where $\Sigma^{IJ}=\half\,\epsilon^{IJ}{}_{KL}e^K\wedge e^L$, $A_{IJ}$ is a Lorentz $SO(3,1)$ connection 
and $F_{IJ}$ is a curvature two-form corresponding to the connection given by
$F_{IJ}=dA_{IJ}+A_{IK}\wedge A^{K}~_{J}$. Our strategy shall be to construct the symplectic structure for the action given in eqn. \eqref{lagrangian1}.  The symplectic potential is obtained to be \cite{Chatterjee:2008if}:
\begin{equation}\label{theta_delta}
16\pi G\gamma~\Theta(\delta)=\gamma~\delta\Sigma_{IJ}\wedge A^{IJ}-\delta(e_{I}\wedge e_{J})\wedge A^{IJ}=-2~\delta(e^{I}\wedge e^{J})\wedge A^{(H)}_{IJ},
\end{equation}
where $A^{(H)}_{IJ}=(1/2)[\,A_{IJ}-(1/2)\,\epsilon_{IJKL}\,A^{KL}]$. From the symplectic one-form eqn. \eqref{theta_delta},  one then constructs the symplectic current $\mathcal{J}(\delta_{1}, \delta_{2})=\delta_{1}\Theta(\delta_{2})- \delta_{2}\Theta(\delta_{1})$ which is closed on-shell. The resulting symplectic current is
\begin{equation}\label{jdelta}
\mathcal{J}\left( \delta_1, \delta_2
\right)=\dfrac{1}{8\pi G\gamma}\left\lbrace \delta_{[1}\left( e_{1}\wedge
e_{2}\right) \right\rbrace \wedge\left\lbrace \delta_{2]}\left(
A_{IJ}-\frac{\gamma}{2} \epsilon_{IJ}{}^{KL}A_{KL}\right)\right\rbrace.
\end{equation}
The symplectic structure is obtained from the symplectic current, eqn. \eqref{jdelta}
 in a similar way described previously in this section. This gives us \cite{Chatterjee:2008if}:
\begin{eqnarray}\label{Palatini_1}
\Omega(\delta_{1}, \delta_{2}
)&=&\frac{1}{8\pi G\gamma}\int_{M}\left[ \delta_{1}(e^{I}\wedge
e^{J})~\wedge\delta_{2}A^{(H)}_{IJ} -\delta_{2}(e^{I}\wedge
e^{J})~\wedge\delta_{1}A^{(H)}_{IJ} \right] \nn
~~&&~~~~~~~~~~~~+\frac{1}{8\pi G\gamma}\int_{S_{\Delta}}\left[
\delta_{1}{}^2\mbf{\epsilon}~\delta_{2}\{\mu_{(m)}+ \gamma\psi_{(\l)}\} -
\delta_{2}{}^2\mbf{\epsilon}~\delta_{1}\{\mu_{(m)} + \gamma\psi_{(\l)}\}\right].
\end{eqnarray}
The function $\psi_{(\ell)}$ is a potential for the surface gravity $\kappa_{(\ell)}$ and is defined by $\lie_\ell\psi_{(\ell)}=\kappa_{(\ell)}$.
Similarly, $\mu_{(m)}$ is the potential for $i(\epsilon-\bar\epsilon)$, and is defined by $\lie_\ell\mu_{(m)}=i(\epsilon-\bar\epsilon)$.
The quantity ${}^2\epsilon$ is the area two form on the spherical cross sections $S_{\Delta}$ of the horizon. The fields 
$\psi_{(\ell)}$ and $\mu_{(m)}$ are assumed to satisfy the boundary condition that $\psi_{(\ell)}=0$ and $\mu_{(m)}=0$ at some initial cross section of the horizon.

%
%

%
\section{Lorentz Transformations and Hamiltonian charges on $\Delta$}
To find the charges arising due to 
local Lorentz transformations,  we take a local basis consisting of the co-tetrads $e^I$. 
The co-tetrads and the connection transform under a Lorentz transformation in the following way.
\beq
e^{I}&\rightarrow &\Lambda^{I}{}_{J}\,e^{J}\\
A^{IJ}&\rightarrow& (\Lambda^{-1})^{I}{}_{K}\,A^{KL}\,\Lambda_{L}{}^{J}+ (\Lambda^{-1})^{I}{}_{K}\,d\Lambda^{KJ}
\eeq
where $\Lambda^I~_{J}$ is the Lorentz transformation matrix. The variations of 
the co tetrads and the connection due to infinitesimal Lorentz transformations, $\Lambda^{I}{}_{J}=(\delta^{I}{}_{J}+\varepsilon \,\epsilon^{I}{}_{J})$, 
are given by the following: 
\beq
\delta_\epsilon e^I&=&\epsilon^{I}{}_{J}\,e^{J}\label{delta_e}\\
\delta_\epsilon A^{IJ}&=&d\epsilon^{IJ} +A^{IK}\epsilon_{K}{}^{J}+A^{JK}\,\epsilon^{I}{}_{K}\label{delta_a},
\eeq
where $\epsilon ^I_J$ are the generators of the Lorentz transformations as discussed in the Section (II).
We also require the expression for the variation of $\Sigma_{IJ}$ and that of $(e_{I}\wedge e_{J})$. 
After a bit of algebra, one can show that,
\begin{eqnarray}
\delta_\epsilon \Sigma_{IJ}
&=&\epsilon^{K}{}_{I}\,\Sigma_{J}{}^{K}-\epsilon^{K}{}_{J}\,\Sigma_{I}{}^{K},\label{delta_sigma}\\
\delta_\epsilon(e_{I}\wedge e_{J})&=&\epsilon^{I}{}_{K}\,e^{K}\wedge e^{J}+\epsilon^{J}{}_{K}\,e^{I}\wedge e^{K}\label{delta_ee}.
\end{eqnarray}

Let us look at the $\gamma$- independent symplectic structure, also called the Palatini symplectic structure.
The action of the Lorentz transformations on the fields, eqns. \eqref{delta_sigma} and \eqref{delta_a},
in the bulk symplectic structure, eqn. \eqref{Palatini_1} leads to:
\beq\label{omega_variation1}
\Omega_{B}(\delta_\epsilon,\delta)&=-&\frac{1}{16\pi G}\int_{\mathcal{M}}(-\epsilon^K~_J\Sigma_{IK}+\epsilon^K~_I\Sigma_{JK})\wedge\delta A^{IJ}-\delta\Sigma_{IJ}\wedge(d\epsilon^{IJ} +A^{IK}\epsilon_{K}{}^{J}+A^{JK}\,\epsilon^{I}{}_{K}),
\eeq
where the subscript $B$ denotes the bulk part of the symplectic structure. The first and the second terms in the above equation add, and so does the fourth and the fifth term. The third term in
eqn. \eqref{omega_variation1}  may be rewritten as 
\beq
\delta\Sigma_{IJ}\wedge d\epsilon^{IJ}&=&d(\delta\Sigma_{IJ}\,\epsilon^{IJ})-\delta \left(d\Sigma_{IJ}\right)\epsilon^{IJ}\nn
&=&d(\delta\Sigma_{IJ}\,\epsilon^{IJ})+\delta(A_I{}^K\wedge\Sigma_{KJ}+A_J{}^K\wedge\Sigma_{IK})\epsilon^{IJ}.
\eeq
Using these expressions in the symplectic structure eqn. \eqref{omega_variation1}, 
we note that the terms with $\delta \Sigma_{IJ}$ cancel each other while those with $\delta A_{IJ}$ cancel 
for the Lorentz transformations which belong to the symmetry group on a WIH. After some
simple algebra, we obtain the following quantity on the cross- sections of the horizon:
\beq\label{bulk_term_charge_1}
\Omega_{B}(\delta_\epsilon,\delta)&=&\frac{1}{16\pi G}\int_{S_{\Delta}}\delta\Sigma_{IJ}\,\epsilon^{IJ}.
\eeq

Similarly, for the $\gamma$- dependent symplectic structure, also called the Holst term, using the action of
Lorentz transformations on the tetrad eqns. \eqref{delta_ee} and the connection variables \eqref{delta_a}, we get 
the symplectic structure \eqref{Palatini_1}  in the following form:
\begin{eqnarray}\label{omega_variation2}
\Omega_{B}(\delta_\epsilon,\delta)=\frac{1}{8\pi G\gamma }\int_{\mathcal{M}}
[\epsilon^{I}{}_{K}\, (e^{K}\wedge e^{J})\wedge\delta A_{IJ}-\delta\, (e_{I}\wedge e_{J})\wedge A^{I}{}_{K}\epsilon_{K}{}^{J}]- \frac{1}{16\pi G\gamma }\int_{\mathcal{M}}\delta\, (e_{I}\wedge e_{J})\wedge d\epsilon^{IJ}. 
\end{eqnarray}
In the eqn. \eqref{omega_variation2}, the quantity involving $d\epsilon^{IJ} $ may also be rewritten in the following way:
\begin{eqnarray}
\delta (e_{I}\wedge e_{J})\wedge d\epsilon^{IJ}=d[\delta(e_{I}\wedge e_{J})\,\epsilon^{IJ}] +2\, \delta[A_{I}{}^{K}\wedge (e_{K}\wedge e_{J})]\epsilon^{IJ}.
\end{eqnarray}
This simplifies the symplectic structure eqn. \eqref{omega_variation2}, and we get the following:
\begin{eqnarray}
\Omega_{B}(\delta_\epsilon,\delta)&=&\frac{1}{8\pi G\gamma }\int_{\mathcal{M}}
[\epsilon^{I}{}_{K}\, (e^{K}\wedge e^{J})\wedge\delta A_{IJ}-\delta\, (e_{I}\wedge e_{J})\wedge A^{I}{}_{K}\epsilon_{K}{}^{J}]
- \frac{1}{16\pi G\gamma }\int_{\mathcal{M}}d[\delta\, (e_{I}\wedge e_{J})\wedge \epsilon^{IJ} ]\nn
&-&\frac{1}{8\pi G\gamma }\int_{\mathcal{M}}
\delta[A^{I}{}_{K}\wedge\, (e^{K}\wedge e^{J})]\,\epsilon_{IJ}.
\end{eqnarray}
In the above expression, the quantity $\delta\, (e_{I}\wedge e_{J})\wedge A^{I}{}_{K}\epsilon_{K}{}^{J}$ cancel with each other
whereas, the quantity $\epsilon^{I}{}_{K}\, (e^{K}\wedge e^{J})\wedge\delta A_{IJ}$ vanish for all those Lorentz transformations
which belong to the symmetry group on the WIH. The remaining term on the horizon cross- section is:
\begin{eqnarray}\label{bulk_term_charge_2}
\Omega_{B}(\delta_\epsilon,\delta)=- \frac{1}{16\pi G\gamma }\int_{S_{\Delta}}\delta\, (e_{I}\wedge e_{J})\wedge \epsilon^{IJ}.
\end{eqnarray}
So, combining these two equation, eqn. \eqref{bulk_term_charge_1} and eqn. \eqref{bulk_term_charge_2},
we note that for Lorentz transformations belonging to the $ISO(2)\ltimes \mathbb{R}$, the bulk contribution of 
the full Holst action to the symplectic structure is reduced to:
\begin{eqnarray}\label{bulk_Holst_1}
\Omega_{B}(\delta_\epsilon,\delta)=- \frac{1}{16\pi G\gamma }\int_{S_{\Delta}}\delta\, (e_{I}\wedge e_{J}-\gamma \Sigma_{IJ})\wedge \epsilon^{IJ}.
\end{eqnarray}
For $\epsilon_{IJ}=R_{IJ}=2im_{[I}\bar m_{J]}$, the symplectic structure in eqn. \eqref{bulk_Holst_1} 
gives the Hamiltonian charge generating this transformation on 
the phase space. Since this is the only space rotation, we shall denote it by $-J$ and the only contribution comes
through the $\gamma$ dependent symplectic structure:
\begin{eqnarray}
\Omega_{B}(\delta_R,\delta)=- \frac{1}{8\pi G\gamma }\int_{S_{\Delta}}\delta\, {}^{2}\epsilon
=- \delta\left(\frac{\mathcal{A}}{8\pi G\gamma }\right)\equiv\delta(-J).
\end{eqnarray}
So, $(\mathcal{A}/8\pi G\gamma)$ is the generator of rotations on the phase- space of isolated horizons.
For $\epsilon_{IJ}=B_{IJ}=-2l_{[I}n_{J]}$, we shall denote the charge by $K$ as it is a boost on the horizon and the only contribution comes
through the $\gamma$- independent symplectic structure eqn. \eqref{bulk_Holst_1}.
\begin{eqnarray}
\Omega_{B}(\delta_B,\delta)=\frac{1}{8\pi G }\int_{S_{\Delta}}\delta\, {}^{2}\epsilon
=\delta\left(\frac{\mathcal{A}}{8\pi G}\right)\equiv\delta(K).
\end{eqnarray}
Again, $(\mathcal{A}/8\pi G)$ is the generator of boosts on the phase- space of isolated horizons. Two simple
statements of immense importance arises quite simply from here. First is the relation
$K=\gamma J$ which has important implications for quantum gravity and is usually referred to as the linear simplicity constraint
\cite{Rovelli:2013osa}. Secondly, 
the area of the horizon is linked to the angular momentum through $\mathcal{A}=8\pi G \gamma J$. 

Several comments are in order. First, one may have some 
contribution from the surface symplectic structure too. Such terms arise
in the proof of the classical first law of black hole mechanics and is crucial 
for constructing a Hamiltonian function corresponding to the null evolution of the horizon \cite{Chatterjee:2008if}.
However, for Lorentz transformations, contributions from the boundary symplectic structure vanish. In the Appendix (B),
a detailed proof is presented for each of these transformations which belong to the $ISO(2)\ltimes \mathbb{R}$.
Secondly, one may also enquire as to why such charges arise in the first place. For the Lorentz boost,
this should be clear since the boost is actually a global symmetry on the horizon and hence, it is natural to
have a charge generating the boost transformation on the phase- space. Indeed, the claim that horizon area must be canonically conjugate to 
the boost is quite well known \cite{Carlip:1993sa,Massar:1999wg,Wall:2010cj,Wall:2011hj, Chatterjee:2015lwa}. Here, we provide a natural
way to obtain this result. For the Lorentz rotations, the situation is different. It is a local symmetry and hence, it's bulk
Hamiltonian generator must be zero by equations of motion. However, due to the presence of horizons,
it has become a genuine symmetry rather than a pure gauge (This argument may as well be true for the other two generators
$P_{IJ}$ and $Q_{IJ}$ belonging to the Lorentz subgroup, but they are identically zero, which may also
be due to the null nature of the boundary itself.). Interestingly, it is a well known fact that in the presence of boundaries,
local symmetries lead to genuine observables and several examples have been discussed in the literature. For example,
consider the Chern-Simons theory on a $3$- manifold, say a disc $\mathcal{D}\times \mathbb{R}$, with $\mathbb{R}$
playing the role of time. In this case, the gauge transformations take field configurations
in the bulk to their gauge equivalent ones, but on the boundary, they become global symmetries \cite{witten}. 
In gravity too, it is well known that diffeomorphisms are broken on the boundary (see \cite{Szabados} for a detail review). 
The gauge motions
due to diffeomorphisms relate gauge equivalent geometries in the bulk, but they give rise to observables
on the boundary. The very well known edge states of gauge theories are examples of such kind.

Let us summarise the findings of this section. We have determined the
effect of Lorentz transformations 
on the spacetime fields and obtained the Hamiltonian charges
due to the boost and the rotation subgroups of the little group of the Lorentz group. 
Notably, both of these transformations are generated 
by quantities related to the area of the horizon. This
expression shall become useful for developing a the quantum theory of the horizon.


%
\section{Non- minimal scalar couplings to gravity}
The effect of non- minimal scalar couplings to gravity may be expressed in a simple way through the
first order Holst action by inclusion of some simple modifications \cite{Chatterjee:2009vd}. The non- minimal scalar coupled
Holst action is given by:
\begin{eqnarray}\label{lagrangian2}
-16\pi G\gamma~L &=& \gamma\, f(\phi)\,\Sigma_{IJ}\wedge F^{IJ}~-~ f(\phi)\,e_{I}\wedge e_{J}\wedge F^{IJ}~-~\gamma~d\{f(\phi)\,\Sigma_{IJ}\wedge A^{IJ}\} \nonumber\\
&& +~d\{ f(\phi)\,e_{I}\wedge e_{J}\wedge A^{IJ}\}- 8\pi GK(\phi)^{*}d\phi\wedge d\phi +16\pi GV(\phi)\epsilon,
\end{eqnarray}
where $V(\phi)$ is a potential for the scalar field, $\epsilon$ is the $4$- dimensional volume,  and the quantity $K(\phi)$ is a scalar function 
given by:
\begin{equation}
K(\phi)=\left[1+(3/16\pi G)\left(f^{\prime 2}(\phi)/f(\phi)\right)\right].
\end{equation}
The symplectic structure corresponding to this action eqn \eqref{lagrangian2} is
obtained to be the following (see \cite{Chatterjee:2009vd} for a detail derivation):
\begin{eqnarray}\label{Palatini_2}
\Omega(\delta_{1}, \delta_{2}
)&=&\frac{1}{8\pi G\gamma}\int_{M}\left[ \delta_{1}\{ f(\phi)\,e^{I}\wedge
e^{J})\}~\wedge\delta_{2}A^{(H)}_{IJ} -\delta_{2}\{f(\phi)\,e^{I}\wedge
e^{J}\}~\wedge\delta_{1}A^{(H)}_{IJ} \right\} \nn
&&~~~~~+\frac{1}{8\pi G\gamma}\int_{S_{\Delta}}\left[
\delta_{1}\{ f(\phi){}^2\mbf{\epsilon}\}~\delta_{2}\{\mu_{(m)}+ \gamma\psi_{(\l)}\} -
\delta_{2}\{ f(\phi){}^2\mbf{\epsilon}\}~\delta_{1}\{\mu_{(m)} + \gamma\psi_{(\l)}\}\right]\nonumber\\
&&~~~~~~~~~~~~~~~~~~~~~~+\int_{M}\,K(\phi)\left[\delta_{1}({}^{*}d\phi)\delta_{2}\phi~-~\delta_{2}({}^{*}d\phi)\delta_{1}\phi\right].
\end{eqnarray}
Using the transformations for the tetrads and the connection variables under Lorentz transformations, we note that
the following boundary contribution remains on the phase- space eqn. \eqref{Palatini_2}, if we restrict
to the Lorentz transformations belonging to the $ISO(2)\ltimes \mathbb{R}$, which is the symmetry group of the WIH: 
\begin{eqnarray}
\Omega_{B}(\delta_\epsilon,\delta)=- \frac{1}{16\pi G\gamma }\int_{S_{\Delta}}\delta\,\left[f(\phi_{0})\,\,e_{I}\wedge e_{J}-\gamma f(\phi_{0})\,\Sigma_{IJ}\,\right]\,\epsilon^{IJ}.
\end{eqnarray}
Note that here, we have fixed the value of the scalar field $\phi=\phi_{0}$ on the horizon. For $\epsilon_{IJ}=R_{IJ}=2im_{[I}\bar m_{J]}$, the symplectic structure gives the Hamiltonian charge generating this transformation on the phase space. This Hamiltonian is 
again the area of the horizon, but now is modified by the scalar field function, $\mathcal{A}\,f(\phi_{0})=8\pi G\gamma J$.
The boost generator is again the area, modified by the scalar field
$\mathcal{A}\,f(\phi_{0})=8\pi G K$, although the simplicity constraint $K=\gamma J$
still holds good. The boundary contribution of the symplectic structure can be shown to be vanishing along the lines of the
Holst symplectic structure.


\section{Entropy of the Weak Isolated Horizon}
On the horizon cross-section, the quantum states are in the representation
of $iso(2)$. In the Appendix, the representation theory has been developed and the eigenstates of the
angular momentum is obtained. As argued in that Appendix C, the representation appropriate for an IH 
corresponds to $p=0$. Since the spacetime algebra is faithfully represented on the phase-space, 
the eigenstates on the angular momentum ($J$) may be used 
to determine the spectrum of the area operator $\mathcal{A}|j\rangle=8\pi G \gamma J\,|j\rangle=8\pi G\hbar \gamma  j\, |j\rangle$. The area eigenvalue $\mathcal{A}$ is then $8\pi G \gamma\hbar j$. This is similar to the result of \cite{Ashtekar:2000eq},
where this arises as a condition on the level of the boundary Chern-Simons theory, and
is essential for quantising this topological theory. In the present scenario, this condition arises naturally due to geometry of the WIH formalism.

Let us consider the surface $S_{\Delta}$ tessellated by a number of patches, much like the surface of a soccer ball. The concept of tessellation follows from the representation used for the area or equivalently, the rotation generator $J$ above. A quantum state of area $S_{\Delta}$ is labeled by an integer or half-integer $|J\rangle$ implies the area of each tessellated patch is also labeled by integers or half-integers and $|J\rangle$ is described by a tensor product structure $|J\rangle=\otimes_i|j_{i}\rangle$ where $i$ is the label for tessellated patches. The area operator is taken to be acting on tessellations as follows: $\mathcal{A}=\oplus_i\mathcal{A}_{i}$ where each of the area patch contributes an area $8\pi \gamma \ell_{p}^{2}j_{i}$. Thus, $J=\sum_{i} j_{i}$. This equation is the basis for calculating the black hole entropy which is obtained by determining the number of independent ways the configurations $\{j_i\}$ can be chosen such that for a fixed $J$ the condition $J=\sum_ij_i$ is satisfied. 

However, the choice of independent tessellations is subject to diffeomorphism constraints. Along the same line of arguments used in loop quantum gravity \citep{Ashtekar:2000eq}, we can fix these constraints by coloring the tessellations. However, this process of fixing the diffeomorphism gauge makes the tessellations distinguishable. Suppose in the partition of $J=N/2$, the number $n_i=2j_i$ is shared by $s_i$ tessellations. So $\sum_is_in_i=N$ and $\sum_is_i$ is the total number of tessellations. So the total number of independent configurations is given by
\begin{equation}
\Omega=\frac{(\sum_is_i)!}{\prod_is_i!}.
\end{equation}
Varying $\log\Omega$ subject to the constraint $\delta\sum_is_in_i=0$ yields the most likely configuration $s_i=(\sum_is_i)\exp(-\lambda n_i)$ where the variation parameter $\lambda$ is to be determined from the constraint $\sum_i\exp(-\lambda n_i)=1$ where $n_i=1,...,N$. This gives $\lambda=\log 2-2^{-N}+o(2^{-2N})$ for large $N$ and entropy $S=\lambda N$. Substituting $N$,
%
%
\begin{equation}\label{entropy}
S=\frac{\mathcal{A}\log 2}{8\pi\gamma\ell_{p}^{2}}+e^{-{\mathcal A}\log 2/8\pi\gamma\ell_p^2},
\end{equation}
and for the choice $\gamma=\ln(2)/2\pi$, the leading order Bekenstein-Hawking result is obtained, but also 
gives an exponentially suppressed corrections to the {\em classical} result.
This is quite a surprisingly new finding that follows directly from this {\em it from bit} formulation of a classical isolated horizon. 
This exponential suppression has in fact been argued to arise in some string computations through non- perturbative corrections 
\cite{Dabholkar:2014ema} although, it has not been obtained in the context of loop quantum gravity.


For the non- minimal couplings, the area
spectrum is obtained from the action of the $J$ operator on its states and hence the 
area operator acts as $f(\phi_{0})\,\mathcal{A}|n\rangle=8\pi G\hbar \gamma n |n\rangle $. This implies that
the classical area in the case of non- minimal couplings is given by $8\pi G\hbar \gamma n/f(\phi_{0})$ and the states are again labelled 
by integers or half integers. The entropy for these black holes will also give the usual area law $S=f(\phi_{0})A/4\ell_{p}^{2}$
with exponential correction terms.


\section{Discussion}
Let us first summarise the results obtained here. Using the WIH phase- space
and the Holst action, we have shown that the the horizon area is the generator 
of both the Lorentz boost on the horizon as well as Lorentz rotations on
two- sphere cross sections of the horizon. These two results also imply that
the linear simplicity constraint holds good for expansion- free and shear- free null surfaces.
For the non- minimal couplings, the area gets modified by the value of the scalar field on the horizon.
While some of these results have been discussed in the literature, they have not been derived
using the WIH formalism. The usefulness lies in the fact that all the considerations are limited to
the horizon geometry only, and no reference to the bulk spacetime is needed. Thus all our proofs appeal
to black hole horizons which may even have time dependent geometrical and matter fields just outside the 
horizon. The local proof carried out here put these results on mathematically sound footing.

Three further remarks are in order. The first is about our demand on distinguishable counting method in Section (VI).
One may argue that elementary excitations must (if seen from a quantum field theory perspective) be indistinguishable.
While this may be correct, there exists arguments in the LQG literature \cite{Ashtekar:2000eq}, where the elementary excitations
(here punctures on a sphere) are distinguishable. In the present framework too we assume this and leave its proof for future works.
The second is that the equidistant area spectrum is quite well known and has been argued for quite some time cite
\cite{Bekenstein_Mukhanov, hod, dreyer, Polychronakos_1}. In the framework
of nonperturbative quantum gravity framework too, it has been argued that the area spectrum may indeed become equidistant \cite{Alekseev}.
Our calculation on the other hand used only the classical gravity to reach a similar conclusion.  
Thirdly, the entropy in equation \eqref{entropy} shows a remarkable behaviour: It has the usual 
area law, but is then suppressed exponentially by the area term. This needs to be looked into further
to understand the origin of such non- perturbative terms. Fourthly, the generators $P$ and $Q$
(see \eqref{iso2}) have zero Hamiltonian charges. In the quantum theory, they are related to the raising and lowering operators (see Appendix C).
It is thus natural that they are vanishing for a null surface, since there is no addition or subtraction of area quanta for a isolated black hole horizon.
In a dynamical horizon framework however, one would require such operators and they will play an important role in understanding the formalism
of Hawking radiation.

 
\section*{Appendix}

\subsection{Algebra of Hamiltonian Charges on the Phase- Space}
Let us now check that the algebra of vector fields on the spacetime is faithfully 
represented through the algebra of the corresponding charges on the phase- space.
As is well known several examples exist where one the algebra of the Hamiltonian charges gets an additional extension.
These kind of extensions lead to interesting contribution to the black hole entropy.

For the $\gamma$ independent part of the Holst action, the symplectic structure for the transformation generated by the $\epsilon_{IJ}$ leads 
to (see eqn. \eqref{bulk_term_charge_1}):
\begin{eqnarray}\label{eq1}
\Omega_{B}(\delta_{\epsilon},\delta)&=&\frac{1}{16\pi G}\int \delta\Sigma_{IJ}\,\epsilon^{IJ},
\end{eqnarray}
where $\delta$ is any arbitrary vector field on the phase space. For the  second transformation generated by 
$\eta_{IJ}$ as given in eqn. \eqref{delta_sigma}, the previous expression in eqn. \eqref{eq1}, simplifies to 
\begin{eqnarray}\label{eq2}
\Omega_{B}(\delta_{\epsilon},\delta_{\eta})&=&\frac{1}{16\pi G}\int \left(-\eta^{K}{}_{J}\Sigma_{IK}+\eta^{K}{}_{I}\Sigma_{JK}\right)\,\epsilon^{IJ}.
\end{eqnarray}
Let us take the following simple generators, $\epsilon_{IJ}=-2l_{[I}n_{J]}$ and $\eta_{IJ}=2m_{[I}l_{J]}+2\bar m_{[I}l_{J]}$. This leads 
to the following expression using eqn. \eqref{eq2} :
\begin{eqnarray}\label{eq3}
\Omega_{B}(\delta_{\epsilon},\delta_{\eta})=\frac{1}{8\pi G}\int \left(\Sigma_{m\ell}+\Sigma_{\bar m\ell}\right),
\end{eqnarray}
where $\Sigma_{m\ell}=\Sigma_{IJ}\,m^{I}\,\ell^{J}$ and $\Sigma_{\bar{m}\ell}=\Sigma_{IJ}\,\bar{m}^{I}\,\ell^{J}$
are the two shorthand notations for these expressions.
Also, for $\epsilon_{IJ}=2m_{[I}l_{J]}+2\bar m_{[I}l_{J]}$, the eqn.\eqref{eq1} gives
\begin{eqnarray}\label{eq4}
\Omega_{B}(\delta_{\epsilon},\delta)&=&\frac{1}{16\pi G}\int \delta\Sigma_{IJ}\,\epsilon^{IJ}=\frac{1}{8\pi G}\int \delta\left(\Sigma_{m\ell}+\Sigma_{\bar m\ell}\right)\equiv\delta H_{\epsilon}.
\end{eqnarray}
Note from eqn. \eqref{eq3} and eqn. \eqref{eq4} that
$\Omega_{B}(\delta_{\epsilon},\delta)=\delta\Omega_{B}(\delta_{\epsilon},\delta_{\eta})$. 
This gives $\delta\{H_{\epsilon},H_{\eta}\}=\delta H_{[\epsilon,\eta]}=\delta H_{\epsilon}$ and hence,
it faithfully represents the vector field algebra of $[B,P]=P$ given in eqn. \eqref{iso2}.

Similarly, let us consider the transformation  generated by the rotation
generator $\epsilon_{IJ}=2im_{[I}\bar m_{J]}$ and that due to $\eta_{IJ}=2m_{[I}l_{J]}+2\bar m_{[I}l_{J]}$.
Following the similar method as above, we get from eqn. \eqref{eq2} that :
\begin{eqnarray}
\Omega_{B}(\delta_{\epsilon},\delta_{\eta})=\frac{i}{8\pi G}\int \left(\Sigma_{m\ell}+\Sigma_{\ell\bar m}\right),
\end{eqnarray}
where $\Sigma_{\ell m}=\Sigma_{IJ}\,\ell^{I}\,m^{J}$ and $\Sigma_{\ell\bar{m}}=\Sigma_{IJ}\,\ell^{I}\,\bar{m}^{J}$.
Also, for $\pi_{IJ}=2im_{[I}l_{J]}-2i\bar m_{[I}l_{J]}$, the symplectic structure, eqn. \eqref{eq1} is
\begin{eqnarray}
\Omega_{B}(\delta_{\pi},\delta)&=&\frac{1}{16\pi G}\int \delta\Sigma_{IJ}\,\pi^{IJ}=\frac{i}{8\pi G}\int \delta\left(\Sigma_{m\ell}+\Sigma_{\ell\bar m}\right)\equiv\delta H_{\pi}.
\end{eqnarray}
This is clearly equal to $\delta\Omega_{B}(\delta_{\epsilon},\delta_{\eta})$. Hence, we get that
$\delta\{H_{\epsilon},H_{\eta}\}=\delta H_{[\epsilon,\eta]}=\delta H_{\pi}$
it faithfully represents the vector field algebra of $[R,P]=Q$ given in eqn. \eqref{iso2}.

This exercise may be extended to the generator $\epsilon_{IJ}=2im_{[I}\bar m_{J]}$ and 
$\eta_{IJ}=2im_{[I}l_{J]}-2i\bar m_{[I}l_{J]}$, for which we get:
\begin{eqnarray}
\Omega_{B}(\delta_{\epsilon},\delta_{\eta})=\frac{-1}{8\pi G}\int \left(\Sigma_{m\ell}+\Sigma_{\bar m \ell}\right).
\end{eqnarray}
Also, for $\pi_{IJ}=2m_{[I}l_{J]}+2\bar m_{[I}l_{J]}$,
\begin{eqnarray}
\Omega_{B}(\delta_{\pi},\delta)&=&\frac{1}{16\pi G}\int \delta\Sigma_{IJ}\,\pi^{IJ}=\frac{1}{8\pi G}\int \delta\left(\Sigma_{m\ell}+\Sigma_{\bar m\ell}\right)\equiv\delta H_{\pi}
\end{eqnarray}
This implies that $\Omega_{B}(\delta_{\pi},\delta)= -\delta\Omega_{B}(\delta_{\epsilon},\delta_{\eta})=-\delta\{H_{\epsilon},H_{\eta}\}
=\delta\{H_{\eta},H_{\epsilon}\}
=\delta H_{[\eta,\epsilon]}=\delta H_{\pi}$. This relation is just the reflection of the internal Lorentz algebra $[R,Q]=-P$ given in eqn. \eqref{iso2}.

For $\epsilon_{IJ}=2im_{[I}\bar m_{J]}$ and $\eta_{IJ}=-2l_{[I}n_{J]}$, we get that the Lorentz charges are given by
$\Omega_{B}(\delta_{\epsilon},\delta_{\eta})=0$.
This may be rewritten as $\delta\{H_{\epsilon},H_{\eta}\}=\delta H_{[\epsilon,\eta]}=0$ and is the charge algebra corresponding
to the vector algebra $[R,B]=0$. The charge algebra may also be obtained for $\epsilon_{IJ}=2m_{[I}l_{J]}+2\bar m_{[I}l_{J]}$ and 
$\eta_{IJ}=2im_{[I}l_{J]}-2i\bar m_{[I}l_{J]}$, in which case, we obtain from eqn. \eqref{eq2}  that
$\Omega_{B}(\delta_{\epsilon},\delta_{\eta})=0$. This again faithfully represents the algebra $[P,Q]=0$.
We may extend the check for the Lorentz boost generator
$\epsilon_{IJ}=-2l_{[I}n_{J]}$ and the generator $\eta_{IJ}=2im_{[I}l_{J]}-2i\bar m_{[I}l_{J]}$, where we get
\begin{eqnarray}
\Omega_{B}(\delta_{\epsilon},\delta_{\eta})=\frac{i}{8\pi G}\int \left(\Sigma_{m\ell}+\Sigma_{\ell\bar m}\right).
\end{eqnarray}
Also, for $\pi_{IJ}=2im_{[I}l_{J]}-2i\bar m_{[I}l_{J]}$, the contribution to the Hamiltonian charge is given by the following:
\begin{eqnarray}
\Omega_{B}(\delta_{\pi},\delta)&=&\frac{1}{16\pi G}\int \delta\Sigma_{IJ}\,\pi^{IJ}=\frac{i}{8\pi G}\int \delta\left(\Sigma_{m\ell}-\Sigma_{\bar m\ell}\right)\equiv\delta H_{\pi}
\end{eqnarray}
This is equal to $\delta\Omega_{B}(\delta_{\epsilon},\delta_{\eta})$ and hence, the Hamiltonian charge algebra
correctly reflects the vector algebra $[B,Q]=Q$, of Lorentz generators.

The same calculation may be repeated for the $\gamma$- dependent part of the symplectic structure in eqn.\eqref{ bulk_term_charge_2}
and the similar results follow there too.
The Hamiltonian charges now are $\gamma$- dependent though the algebra of these charges
is independent of this Immirzi parameter ($\gamma$). The contribution of the symplectic structure for these Lorentz transformations
may be obtained by using the eqns. \eqref{delta_ee} and \eqref{bulk_term_charge_2} and we get
\begin{eqnarray}\label{eq6}
\Omega_{B}(\delta_{\epsilon},\delta)&=&\frac{-1}{16\pi G\gamma}\int \delta\, (e_{I}\wedge e_{J})\,\epsilon^{IJ}\nn
\Omega_{B}(\delta_{\epsilon},\delta_{\eta})&=&\frac{-1}{16\pi G\gamma}\int \left\{-\eta^{K}{}_{J}\,(e_{I}\wedge e_{K})+\eta^{K}{}_{I}\,(e_{J}\wedge e_{K})\right\}\,\epsilon^{IJ}.
\end{eqnarray}
For $\epsilon_{IJ}=-2l_{[I}n_{J]}$ and $\eta_{IJ}=2m_{[I}l_{J]}+2\bar m_{[I}l_{J]}$, the symplectic structure  in eqn. \eqref{eq6} reduces to
the following form:
\begin{eqnarray}
\Omega_{B}(\delta_{\epsilon},\delta_{\eta})=\frac{-1}{8\pi G\gamma}\int \left(e_{m}\wedge e_{\ell} +e_{\bar m}\wedge e_{\ell}\right),
\end{eqnarray}
where $(e_{m}\wedge e_{\ell})=(e_{I}\wedge e_{J})\,m^{I}\,\ell^{J}$. Also, for $\eta_{IJ}=2m_{[I}l_{J]}+2\bar m_{[I}l_{J]}$, the symplectic structure in eqn. \eqref{eq6} gives the Hamiltonian charge:
\begin{eqnarray}
\Omega_{B}(\delta_{\eta},\delta)&=&\frac{-1}{16\pi G\gamma}\int \delta\, (e_{I}\wedge e_{J})\,\eta^{IJ}=\frac{-1}{8\pi G\gamma}\int \delta\left(e_{m}\wedge e_{\ell} +e_{\bar m}\wedge e_{\ell}\right)\equiv\delta H_{\eta}.
\end{eqnarray}
Note that this is equal to the expression $\delta\Omega_{B}(\delta_{\epsilon},\delta_{\eta})$ and hence,
we get that $\delta\{H_{\epsilon},H_{\eta}\}=\delta H_{[\epsilon,\eta]}=\delta H_{\eta}$. This charge algebra is
reflective of the fact that at the the algebra of vector fields $[B,P]=P$ holds. Also note that though the charges are $\gamma$
dependent, this does not show up in the charge algebra.

Let us now consider the charges due to the two transformations generated by $\epsilon_{IJ}=2im_{[I}\bar m_{J]}$ and $\eta_{IJ}=2m_{[I}l_{J]}+2\bar m_{[I}l_{J]}$. The symplectic structure in eqn. \eqref{eq6} reduces to:
\begin{eqnarray}
\Omega_{B}(\delta_{\epsilon},\delta_{\eta})=\frac{-i}{8\pi G\gamma}\int \left(e_{\ell}\wedge e_{\bar m} +e_{m}\wedge e_{\ell}\right).
\end{eqnarray}
This charge is also obtained by the direct application of the $\pi_{IJ}=2im_{[I}l_{J]}-2i\bar m_{[I}l_{J]}$ on the phase- space quantities
and we get that
\begin{eqnarray}
\Omega_{B}(\delta_{\pi},\delta)&=&\frac{-1}{16\pi G\gamma}\int \delta\, (e_{I}\wedge e_{J})\,\pi^{IJ}
=\frac{-i}{8\pi G\gamma}\int \left(e_{\ell}\wedge e_{\bar m} +e_{m}\wedge e_{\ell}\right)\equiv\delta H_{\pi}=\delta\Omega_{B}(\delta_{\epsilon},\delta_{\eta}).
\end{eqnarray}
Hence, this the Hamiltonian charge equivalent of the algebra $[R,P]=Q$. 

The equivalent charge algebra for
the $[R,Q]=-P$ is obtained as follows. First, for $\epsilon_{IJ}=2im_{[I}\bar m_{J]}$ and $\eta_{IJ}=2im_{[I}l_{J]}-2i\bar m_{[I}l_{J]}$, we get
from eqn. \eqref{eq6} that:
\begin{eqnarray}
\Omega_{B}(\delta_{\epsilon},\delta_{\eta})=\frac{-1}{8\pi G\gamma}\int \left(e_{\ell}\wedge e_{\bar m} +e_{\ell}\wedge e_{\bar m}\right).
\end{eqnarray}
Now, the generator of the Lorentz transformation $\pi_{IJ}=2m_{[I}l_{J]}+2\bar m_{[I}l_{J]}$, gives the following charge on the phase- space:
\begin{eqnarray}
\Omega_{B}(\delta_{\eta},\delta)&=&\frac{-1}{16\pi G\gamma}\int \delta\, (e_{I}\wedge e_{J})\,\pi^{IJ}=\frac{1}{8\pi G\gamma}\int \left(e_{\ell}\wedge e_{\bar m} +e_{\ell}\wedge e_{\bar m}\right)=\delta H_{\eta}
\end{eqnarray}
This is equivalent to $-\delta\Omega_{B}(\delta_{\epsilon},\delta_{\eta})=\delta\{H_{\eta},\,H_{\epsilon}\}=\delta H_{[\eta,\epsilon]}$,
and hence, we get the faithful representation of the algebra $[R, Q]=-P$.

The other algebra of charges are obtained similarly. Let us consider $\epsilon_{IJ}=2im_{[I}\bar m_{J]}$ and $\eta_{IJ}=-2l_{[I}n_{J]}$.
The symplectic structure gives $\Omega_{B}(\delta_{\epsilon},\delta_{\eta})=0$ and hence, this gives
$\delta\{H_{\epsilon},H_{\eta}\}=\delta H_{[\epsilon,\eta]}=0$. This is equivalent to the algebra of $[B, R]=0$ on the phase- space.
This is also carried over to the set of transformations
$\epsilon_{IJ}=2m_{[I}l_{J]}+2\bar m_{[I}l_{J]}$ and $\eta_{IJ}=2im_{[I}l_{J]}-2i\bar m_{[I}l_{J]}$, where 
$\Omega_{B}(\delta_{\epsilon},\delta_{\eta})=0$ which gives $\delta\{H_{\epsilon},H_{\eta}\}=\delta H_{[\epsilon,\eta]}=0$,
and is the phase- space realisation of the algebra $[P,Q]=0$. Also,
for the transformations generated by $\epsilon_{IJ}=-2l_{[I}n_{J]}$ and $\eta_{IJ}=2im_{[I}l_{J]}-2i\bar m_{[I}l_{J]}$, we get:
\begin{eqnarray}
\Omega_{B}(\delta_{\epsilon},\delta_{\eta})=\frac{-i}{8\pi G\gamma}\int \left(e_{m}\wedge e_{\ell} -e_{\bar m}\wedge e_{\ell}\right).
\end{eqnarray}
This must be compared to the charge generated by the transformation $\pi_{IJ}=2im_{[I}l_{J]}-2i\bar m_{[I}l_{J]}$, which gives us
\begin{eqnarray}
\Omega_{B}(\delta_{\eta},\delta)&=&\frac{-1}{16\pi G\gamma}\int \delta\, (e_{I}\wedge e_{J})\,\pi^{IJ}=\frac{-i}{8\pi G\gamma}\int \delta\left(e_{m}\wedge e_{\ell} -e_{\bar m}\wedge e_{\ell}\right)\equiv\delta H_{\pi}.
\end{eqnarray}
This is same as $\delta\Omega_{B}(\delta_{\epsilon},\delta_{\eta})=\delta H_{[\epsilon,\eta]}$ and hence
is same as $[B,Q]=Q$. Thus, all the algebra of charges is exactly that of the algebra of vector fields and hence, the algebra of vector fields is faithfully represented by the algebra of charges on the phase space.

\subsection{The contributions from boundary symplectic structure}
The purpose of this section is to show that the contribution from the
boundary symplectic structure in eqn. \eqref{Palatini_1} vanishes. There are two contribution from the boundary symplectic structure. 
The first one is related to the phase- space scalar function $\psi_{\ell}$.
Note that $d\psi_{(\ell)}=-\kappa_{(\ell)}n+\alpha_{\ell}m +\bar\alpha_{\ell}\bar m$,
where $\alpha_{\ell}$ and $\bar\alpha_{\ell}$ are some scalars.
The variation due to the Lorentz transformations should affect all these quantities. For each of these transformations, let us
obtain these changes one by one.

Let us first consider the transformation due to the boost generator on $\Delta$ given by 
$\eta^{I}{}_{J}= -(\ell^{I}n_{J}-\ell_{J}n^{I})$. Note that on
the horizon $(\ell^{I}, n^{I}, m^{I},\bar m^{I})$ are all fixed. Therefore, the effect 
of transformation on $n_{a}$ is determined as follows:
$\delta_{\eta}n_{a}=\delta_{\eta}e^{I}_{a}n_{I})=\delta_{\eta}(e^{I}_{a})n_{I}=\epsilon^{I}{}_{J}e^{J}_{a}n_{I}=-(\ell^{I}n_{J}-\ell_{J}n^{I})e^{J}_{a}n_{I}=n_{J}e^{J}_{a}=n_{a}$. Similarly, the vector $\ell_{a}$ shall also change
$\delta_{\eta}\ell_{a}=\delta_{\eta}(e^{I}_{a}\ell_{I})=\delta_{\eta}(e^{I}_{a})\ell_{I}=\epsilon^{I}{}_{J}e^{J}_{a}\ell_{I}=-(\ell^{I}n_{J}-\ell_{J}n^{I})e^{J}_{a}\ell_{I}=-\ell_{J}e^{J}_{a}=-\ell_{a}$. This expression is consistent with the fact that $\ell^{a}\,n_{a}=-1$. 
For these transformations, one also obtains that $\delta_{\eta}m_{a}=0$ as well as $\delta_{\eta}\bar m_{a}=0$.
These results now may be used to obtain the variation for the surface gravity $\kappa_{(\ell)}$ due to the transformation
given above. Since $\ell^{a}\nabla_{a}\ell^{b}=\kappa_{(\ell)}\ell^{b}$,  the variation
is $\delta_{\eta}\{\ell^{a}\nabla_{a}\ell^{b}=\kappa_{(\ell)}\ell^{b}\}$ which leads to the relation
$-\ell^{a}\nabla_{a}\ell^{b}-\ell^{a}\nabla_{a}\ell^{b}=\delta_{\eta}\{\kappa_{(\ell)}\}\ell^{b} -\kappa_{(\ell)}\ell^{b}$, 
and hence one gets that $\delta_{\eta}\,\kappa_{(\ell)}=-\kappa_{(\ell)}$. Therefore, if one looks into
the expression of $d\psi_{(\ell)}$, the variation of the first term is 
$\delta_{\eta}(\kappa_{(\ell)} n_{a})=\delta_{\eta}\,\{\kappa_{(\ell)}\} n_{a}+\kappa_{(\ell)} \delta_{\eta}\,n_{a}=-\kappa_{(\ell)} n_{a}+\kappa_{(\ell)} n_{a}=0$.
So, $\delta_{\eta}d\psi_{(\ell)}=\delta(\alpha_{\ell})m +\delta(\bar\alpha_{\ell})\bar m$, and
hence $\lie_{\ell}\delta_{\eta}\psi_{(\ell)}=0$ and hence $\delta_{\eta}\psi_{(\ell)}$ depends
only on the coordinates of the two- sphere. If we set
$\delta_{\eta}\psi_{(\ell)}=0$ at some initial cross- section, it is going to remain the same throughout the horizon.
Also, due to the expressions of variation of $m_{a}$ and $\bar{m}_{a}$, we get that $\delta_{\eta}{}^{2}\epsilon=0$.

A similar logic also applies for the transformation $\eta^{I}{}_{J}= i(m^{I}\bar m_{J}-\bar m_{J}m^{I})$. We get that
$\delta_{\eta}n_{a}=0$ and $\delta_{\eta}\ell_{a}=0$ and furthermore,
$\delta_{\eta} m_{a}=-im_{a}$ and $\delta_{\eta} \bar m_{a}=i\bar m_{a}$.
These transformations obviously imply that the variation
of surface gravity $\delta_{\eta}\kappa_{(\ell)}=0$ and the
variation of the scalar $\psi_{(\ell)}$ is given by 
$\delta_{\eta}d\psi=\delta(\alpha_{\ell})m -i(\alpha_{\ell})m+\delta(\bar\alpha_{\ell})\bar m+i(\bar\alpha_{\ell})m$.
Hence, this immediately gives that $\lie_{\ell}\delta_{\eta}\psi=0$. Again, if we set
$\delta_{\eta}\psi_{(\ell)}=0$ at some initial cross- section,  it shall remain vanishing on the horizon. 
Additionally, these variations also lead to $\delta_{\eta}{}^{2}\epsilon=0$.

The next set of transformation is $\eta^{I}{}_{J}= (m^{I}\ell_{J}-m_{J}\ell^{I})+(\bar m^{I}\ell_{J}-\bar m_{J}\ell^{I})$. 
We get the following: $\delta_{\eta}n_{a}=(m_{a}+\bar m_{a})$, $\delta_{\eta}\ell_{a}=0$,
$\delta_{\eta} m_{a}=\ell_{a}$ and $\delta_{\eta} \bar m_{a}=\ell_{a}$. Naturally, this
leads to $\delta_{\eta}\kappa_{(\ell)}=0$ and so, the complete variation
is due to $\delta_{\eta}d\psi=-\kappa_{(\ell)} (m_{a}+\bar m_{a})+\delta(\alpha_{\ell})m +\delta(\bar\alpha_{\ell})\bar m$, and 
this gives us $\lie_{\ell}\delta_{\eta}\psi_{(\ell)}=0$. If we set
$\delta_{\eta}{}^{2}\epsilon=0$  at some initial cross- section, it shall remain vanishing throughout
the horizon. Also $\delta_{\eta}{}^{2}\epsilon=0$ when pulled back to the horizon. For 
the transformation $\eta^{I}{}_{J}= i(m^{I}\ell_{J}-m_{J}\ell^{I})-i(\bar m^{I}\ell_{J}-\bar m_{J}\ell^{I})$, too, 
$\delta_{\eta}\psi_{(\ell)}$ shall remain vanishing throughout
the horizon when pulled back to the horizon.

The second quantity which arises in the boundary term of the symplectic structure in \eqref{Palatini_1} is 
due to $d\mu_{(m)}=-(\epsilon-\bar\epsilon)n_{a}+(\beta_{m} m_{a}+\bar \beta_{m}\bar m_{a})$, where $\beta_{m}$ and $\bar \beta_{m}$
are two scalars whose exact form is not required, and  $\ell^{a}\nabla_{a}m^{b}=(\epsilon-\bar \epsilon)m^{b}$. Let us now look for the variations due to the first set of transformations $\eta^{I}{}_{J}= -(\ell^{I}n_{J}-\ell_{J}n^{I})$, 
for which we had that $\delta_{\eta}n_{a}=n_{a}$, $\delta_{\eta}\ell_{a}=-\ell_{a}$, 
$\delta_{\eta}m_{a}=0$ and $\delta_{\eta}\bar m_{a}=0$. Now, since the $(\epsilon-\bar \epsilon)$ is determined by
 $\ell^{a}\nabla_{a}m^{b}=(\epsilon-\bar \epsilon)m^{b}$, the variation is  
 $\delta_{\eta}\{\ell^{a}\nabla_{a}m^{b}=(\epsilon-\bar \epsilon)m^{b}\}$ which gives,
$\ell^{a}\nabla_{a}m^{b}=-\delta_{\eta}(\epsilon-\bar \epsilon)m^{b}$ which simplifies to
$\delta(\epsilon-\bar \epsilon)=-(\epsilon-\bar \epsilon)$, and hence $\delta_{\eta}\{-(\epsilon-\bar \epsilon)n_{a}\}=0$.
As a result of this calculations,  $\delta_{\eta}d\mu=\delta(\alpha_{\ell})m +\delta(\bar\alpha_{\ell})\bar m$, and 
therefore, $\lie_{\ell}\delta_{\eta}\mu=0$. If we set
$\delta_{\eta}\mu\mu_{(m)}=0$  at some initial cross- section, it is going to remain the same on
the horizon. One also gets that $\delta_{\eta}{}^{2}\epsilon=0$.

For the transformations $\eta^{I}{}_{J}= i(m^{I}\bar m_{J}-\bar m_{J}m^{I})$, we had already obtained that
$\delta_{\eta}n_{a}=0$, $\delta_{\eta}\ell_{a}=0$,
$\delta_{\eta} m_{a}=-im_{a}$ and $\delta_{\eta} \bar m_{a}=i\bar m_{a}$.
This leads to the condition that $\delta_{\eta}(\epsilon-\bar \epsilon)=0$.
So, $\delta_{\eta}d\mu=\delta(\beta_{m})m -i(\beta_{m})m+\delta(\bar\beta_{m})\bar m+i(\bar\beta_{m})m$, and $\lie_{\ell}\delta_{\eta}\mu=0$. This leads again to $\delta_{\eta}\mu\mu_{(m)}=0$ on the horizon. For these transformation,
one also obtains $\delta_{\eta}{}^{2}\epsilon=0$.

Let us now look at the set of transformations
$\eta^{I}{}_{J}= (m^{I}\ell_{J}-m_{J}\ell^{I})+(\bar m^{I}\ell_{J}-\bar m_{J}\ell^{I})$, for which,
$\delta_{\eta}n_{a}=(m_{a}+\bar m_{a})$,  $\delta_{\eta}\ell_{a}=0$,
$\delta_{\eta} m_{a}=\ell_{a}$ and $\delta_{\eta} \bar m_{a}=\ell_{a}$.
From these transformations, it arises that $\ell^{a}\nabla_{a}m^{b}=(\epsilon-\bar \epsilon)m^{b}$
and therefore, $\delta\{\ell^{a}\nabla_{a}m^{b}=(\epsilon-
\bar \epsilon)m^{b}\}$ which gives
$\ell^{a}\nabla_{a}\ell^{b}=\delta_{\eta}(\epsilon-
\bar \epsilon)m^{b}+(\epsilon-
\bar \epsilon)\ell^{b}$. A simple calculation gives $\delta(\epsilon-
\bar \epsilon)=\{\kappa_{(\ell)} \ell^{b}-(\epsilon-\bar \epsilon)\ell^{b}\}\bar m_{b}=0$.
From these results, we get that $\delta_{\eta}d\mu_{(m)}=-(\epsilon-\bar \epsilon)(m_{a}+\bar m_{a})+\delta(\beta_{m})m +\delta(\bar\beta_{m})\bar m$, and 
$\lie_{\ell}\delta_{\eta}\psi=0$. Thus, $\delta_{\eta}\mu_{(m)}=0$ on the horizon and
$\delta_{\eta}{}^{2}\epsilon=0$, when pulled back to the horizon.
Similarly, for $\eta^{I}{}_{J}= (m^{I}\ell_{J}-m_{J}\ell^{I})+(\bar m^{I}\ell_{J}-\bar m_{J}\ell^{I})$, we get that $\delta_{\eta}\,\mu_{(m)}=0$
and $\delta_{\eta}{}^{2}\epsilon=0$ when pulled back to the horizon.

The similar calculation may also be carried out for the symplectic structure of the non- minimally coupled scalar field 
given in eqn. \eqref{Palatini_2}, and here also, the boundary contributions are zero. Only the bulk term contributes to Hamiltonian.

\subsection{The algebra of $iso(2)$ and its representation}
The transformations corresponding to the the group of $ISO(2)$
is that of inhomogeneous rotations in two dimensional plane given by:
\begin{equation}
x^{1\,\,\prime}=x^{1}\cos\theta -x^{2}\sin\theta +t^{1},~~
x^{2\,\,\prime}=x^{1}\sin\theta +x^{2}\cos\theta +t^{2}.
\end{equation}
This set of transformations may be written as:
\begin{equation}
x^{\mu}\rightarrow x^{\mu\,\,\prime}=G^{\mu}_{\nu}(\theta,t^{\lambda})\,x^{\nu}, ~~~\mu,\nu,\lambda \dots =1,2 ,
\end{equation}
where, the group element for transformation is written in the matrix representation:
\[
G^{\mu}_{\nu}(\theta,t^{\lambda})= 
\begin{bmatrix}
 \cos\theta     & -\sin\theta   & t^{1} \\
   \sin\theta     & \cos\theta   & t^{2}   \\
   0                   &      0           & 1
\end{bmatrix},
\]
where $\theta$ is the rotation angle and $t^{1}$ and $t^{2}$ are the translations. Just like the Poincare
group, this is the semi- direct product of rotations and translations.
\begin{equation}
G^{\mu}_{\nu}(\theta,t^{\lambda})\,G^{\nu}_{\sigma}(\theta^{\prime},t^{\prime\,\,\lambda})=G^{\mu}_{\sigma}(\theta^{\prime\prime},t^{\prime\prime\,\,\lambda}),
\end{equation}
where $\theta^{\prime\prime}=\theta+\theta^{\prime}$ and $t^{\prime\prime\,\,\lambda}=R^{\lambda}{}_{\sigma}\,(\theta)t^{\sigma}
+t^{\prime\lambda}$, with $R^{\lambda}{}_{\sigma}(\theta)=G^{\lambda}{}_{\sigma}(\theta,0)$ being the rotation part of the group transformation. The infinitesimal group elements for rotation ($G_{R}$), and the two translations ($G_{T1}$ and $G_{T2}$) are as follows:
\[
G_{R}(\epsilon,0)= 
\begin{bmatrix}
 1-\epsilon^{2}/2    & -\epsilon   & 0 \\
   \epsilon     &  1-\epsilon^{2}/2   & 0   \\
   0                   &      0           & 1
\end{bmatrix},
\,\,\, G_{T1}(\epsilon,0)= 
\begin{bmatrix}
 1    & 0  & \epsilon \\
  0   &  1   & 0   \\
   0   & 0   & 1
\end{bmatrix},
\,\, \, G_{T2}(\epsilon,0)= 
\begin{bmatrix}
 1    & 0  & 0\\
  0   &  1   & \epsilon    \\
   0   & 0   & 1
\end{bmatrix}.
\]
If we consider these infinitesimal generators to act on the space of functions,
they may be represented as rotations in the $x-y$ plane and translation in the 
$x$ and $y$ directions respectively.
\begin{equation}
J=x(\partial/\partial y) -y(\partial/\partial x), \,\,\, P=(\partial/\partial x),\,\,\, Q=(\partial/\partial y). 
\end{equation}
For these transformations in the matrix form, it is simple to check that the following relations hold:
\begin{eqnarray}
G_{R}(\epsilon,0)G_{T1}(0,\epsilon)-G_{T1}(0,\epsilon)G_{R}(\epsilon,0)&=&G_{T2}(0,\epsilon^{2})-G(0,0),\\
G_{R}(\epsilon,0)G_{T2}(0,\epsilon)-G_{T2}(0,\epsilon)G_{R}(\epsilon,0)&=&-G_{T1}(0,\epsilon^{2})+G(0,0),\\
G_{T1}(0,\epsilon)G_{T2}(0,\epsilon)-G_{T2}(0,\epsilon)G_{T1}(0,\epsilon)&=&0.
\end{eqnarray}
The group representations are obtained by the exponential mapping and on the quantum
states, the action is carried out through $\mathcal{D}(\theta, t^{1}, t^{2})=\exp\{-iJ\theta/\hbar-it_{1}P/\hbar-it_{2}Q/\hbar\}$.
This group representation should satisfy the infinitesimal version given above.
\begin{eqnarray}
\mathcal{D}_{R}(\epsilon,0)\mathcal{D}_{T1}(0,\epsilon)-\mathcal{D}_{T1}(0,\epsilon)\mathcal{D}_{R}(\epsilon,0)&=&\mathcal{D}_{T2}(0,\epsilon^{2})-\mathcal{D}(0,0),\\
\mathcal{D}_{R}(\epsilon,0)\mathcal{D}_{T2}(0,\epsilon)-\mathcal{D}_{T2}(0,\epsilon)\mathcal{D}_{R}(\epsilon,0)&=&-\mathcal{D}_{T1}(0,\epsilon^{2})+\mathcal{D}(0,0),\\
\mathcal{D}_{T1}(0,\epsilon)\mathcal{D}_{T2}(0,\epsilon)-\mathcal{D}_{T2}(0,\epsilon)\mathcal{D}_{T1}(0,\epsilon)&=&0.
\end{eqnarray}
Keeping terms upto $\epsilon^{2}$, this gives us the following algebra of the generators:
\begin{equation}
[\,J,\,P\,]=i\hbar Q, ~~ [\,J,\,Q\,]=-i\hbar P~~ [\,P,\,Q\,]=0.
\end{equation}
One may define a linear combination of these operators to construct the shift operators as follows:
$P_{+}=P+iQ$ and $P_{-}=P-iQ$. This gives $[\,J,P_{+}\,]=\hbar P_{+}$ and $[\,J,P_{-}\,]=-\hbar P_{-}$.
In coordinate representation ($x=r\cos\phi$ and $y=r\sin\phi$), these operators are given by:
$J=(-i\hbar)(\partial/\partial\phi)$, $P_{+}=(-i\hbar)\,e^{i\phi}\left[\partial/\partial r +(i/r)(\partial/\partial \phi)\right]$,
and $P_{-}=(-i\hbar)\,e^{-i\phi}\left[-\partial/\partial r +(i/r)(\partial/\partial \phi)\right]$. 

Note that if we define
an operator $P^{2}=P_{-}P_{+}=P_{-}P_{+}=P^{2}+Q^{2}$, this operator commutes with all the generators
of the algebra, $[P^{2}, J]=0$ and $[P^{2}, P_{+}]=0=[P^{2}, P_{-}]$. So, the states are labelled by
the eigenvalues of $P^{2}$ and $J$. Let us call them $p^{2}$ and $n$ respectively, and the states be labeled by $|p^{2},n\rangle$. We must have $p^{2}\geqslant 0$. 
Let us consider the case $p > 0$ first.
The eigenkets of $J$ with integer eigenvalues are $J|p^2,n\rangle=n\hbar|p^2,n\rangle$.
These kets have a function representation and is given by $(-i\hbar)(\partial/\partial\phi)\langle r,\phi |p^2, n\rangle=n\hbar\langle r,\phi |p^2,n \rangle$ and hence $\langle r,\phi |p^2, n\rangle =f_{n,p}(r)e^{in\phi}$ (that $n$ is an integer follows from the periodicity of the eigenfunctions). Now, a simple algebra shows that $P_{+}$ and $P_{-}$ are
the raising and lowering operators respectively. More precisely,
$P_{+}|p^2, n\rangle=\hbar |p^2,n+1\rangle$ and $P_{-}|p^2, n\rangle=\hbar |p^2,n-1\rangle$.
This gives us that $P_{+} P_{-}|p^2, n\rangle=(\hbar)^{2}\,|p^2,n\rangle$ and therefore, a representation in terms of functions is obtained through
$P_{+} P_{-}\langle r,\phi|p^2, n\rangle=(\hbar)^2\langle r,\phi |p^2,n\rangle$. This gives Bessel's equation for the function $f_{n,p}(r)$:
\begin{equation}\label{B_eqn}
f_{n,p}^{\prime\prime}+(1/r)f_{n,p}^{\prime}+(p^2-n^{2}/r^{2})f_{n,p}(r)=0.
\end{equation}
The modes $n$ correspond to the vibrational modes of a membrane. 
This representation is clearly infinite dimensional and is related to the fact that we have taken the eigenvalues of $P^{2}$ to be nonzero.

Let us now consider the case $p^{2}= 0$. In that case, the eigenvalues of $P$, $Q$ and $P^{2}$ all
vanish, $P_{+} P_{-}| n\rangle=0$. So, the $P_{+}$ and the $P_{-}$ operators do not raise or lower the states but they are used to choose the \emph{physical states}. Indeed, for solutions belonging to the WIH phase-space both $P_{+}$ and $P_{-}$ vanish. In other words, the label $n$ of the states are not raised or lowered on a WIH. This is not unexpected since $n$ gives the total area and the area of WIH does not increase or decrease. Instead, the operators $P_\pm$ must be interpreted
as constraints acting on physical states: $P_{+} |n\rangle=P_{-}| n\rangle=0$. 
Next, because these horizon states have $p^{2}=0$, they must be labeled by $n$ \emph{only}. Let us prove this statement:
From the algebra, it is clear that $P$ and $Q$ (or $P_{+}$ and $P_{-}$) are vector operators, they transform
under rotations: $\{\exp(-iJ\theta)\,P_{\mu}\,\exp(iJ\theta)\}=R_{\mu}{}^{\nu}\,(\theta)\,P_{\nu}$, (here, $\mu,\nu=1,2$
and $P_{1}\equiv P$ and $P_{2}=Q$). So,
if $P$ and $Q$ (or $P_{+}$ and $P_{-}$) have one non- zero eigenvalue, one may use the continuum values of $\theta$ to obtain infinite 
number of non- zero eigenvalues and hence, infinite continuum of states labeled by a continuous degree of freedom $\theta$.
Since $p=0$, the physical states must be labelled only by $n$. This proves our statement. Hence,
the irreducible representation for this case is 1-dimensional and the states are labeled by the integers or half- integers $n$
\cite{Weinberg}. Note that, since $J$ is related to area in this paper, its eigenvalues (which are interpreted as the area quanta), are be taken to be strictly positive, $n>0$.


\subsection*{Acknowedgements}
The authors thank Avirup Ghosh for discussions. AC thanks IUCAA for a visit where parts of this work was carried out.
%
%

\end{document}